\documentstyle[aps,preprint]{revtex}
\begin{document}
\draft
\title{New Sum Rules for an Electron in a Periodic or Nearly 
Periodic Potential} 
\author{Michael R. Geller}
\address{Department of Physics, Simon Fraser University, 
Burnaby, B.C. V5A 1S6 Canada}
\date{\today}
\maketitle

\begin{abstract}
I derive new sum rules for the electronic oscillator strengths
in a periodic or nearly periodic potential, which apply within a single
energy band and between any two bands. 
The physical origin of these sum 
rules is quite unlike that of conventional sum rules,  
and is shown to be
associated with 
a solid-state counterpart to the principle of spectroscopic stability
known in atomic physics.
The new sum rules have important implications for the optical properties
of semiconductor nanostructures, and for the spectroscopy of solids
in general.
\end{abstract}


Bloch's seminal analysis of the quantum 
mechanical motion of a particle in a periodic potential
lies at the heart of the modern electron theory of solids, 
and provides a basis---band theory---for the
entire conceptual framework of the subject. 
The present work is motivated by the recent interest
in the electronic properties of nanometer-scale 
semiconductor structures \cite{Bastard etal},
and focuses on the quantum mechanics of electrons in 
{\it nearly periodic potentials}.
The nearly periodic potentials arise either from 
uniform crystals in the presence of scalar
potentials varying slowly in space (for example, from impurities or gates)  
or from compositionally {\it graded}
crystals with or without applied potentials. 
A general theory of 
stationary states in these nearly periodic potentials
has been recently developed \cite{Geller and Kohn}. 

Motivated by the tremendous interest in the optical 
properties of electrons in semiconductor
nanostructures such as quantum wells, quantum wires,
and quantum dots, we turn here to
the dynamics of electrons in nearly 
periodic potentials. 
First, we shall derive new sum rules for the electronic 
oscillator strengths in a periodic or
nearly periodic potential,
which apply within a single band or between any two bands. 
The intraband and interband sum rules govern the 
optical transition strengths between a given state and all the other
states in the same and different bands, respectively.
Like other known sum rules \cite{Mahan}, they are expected to be extremely
useful both for supplementing microscopic calculations of the optical
properties of solids, and for interpreting
their measurement.
Next, we show that the physical origin of these sum rules
is associated with a particular stability of the optical
absorption spectrum of a Bloch electron in the presence of
perturbations varying slowly on the scale of the lattice 
constant, which reflects the weakness of the resultant 
{\it interband} mixing. This {\it spectroscopic stability} has
a known counterpart in the theory of atomic spectra.
As an application of our results, we shall calculate the optical 
absorption spectrum of an electron in 
a parabolic quantum well, and we examine
the effects of the position-dependent band 
structure on the intraband oscillator strengths
and selection rules.

We begin by deriving an expression for the
dynamic conductivity tensor, 
$\sigma^{ij}(\omega)$, for an electron in a
uniform or graded crystal.
Let $\epsilon_n({\bf k},c)$ denote the energy bands of a uniform alloy 
${\rm A_c B_{1-c}}$, which we assume are known. 
We shall consider a graded crystal 
with a composition $c({\bf r})$ varying slowly
on the scale of a characteristic lattice constant $b$, in the 
presence of a slowly varying scalar 
potential $U({\bf r})$, and we shall for simplicity 
limit ourselves to simple bands. 
The stationary states,
\begin{equation}
 \psi_{n \alpha}({\bf r}) 
= \sum_{\bf l} \Phi^{\alpha}_{n {\bf l}}  
\ a_{n {\bf l}}({\bf r}-{\bf l}) ,
\label{expansion}
\end{equation}
and energies, $\epsilon_{n \alpha}$,
of an electron in a graded crystal may be obtained by 
solving an eigenvalue problem,
\begin{equation}
\sum_{{\bf l}'} {\cal H}^n_{{\bf l l}'} \ \Phi_{n {\bf l}'}^\alpha
+ U({\bf l}) \ \Phi_{n{\bf l}}^\alpha
= \epsilon_{n \alpha} \ \Phi_{n{\bf l}}^\alpha ,
\label{eigenvalue problem}
\end{equation}
with an effective Hamiltonian
\begin{equation}
{\cal H}^n_{{\bf l l}'} \equiv {1 \over v}
\int_{\rm \scriptscriptstyle BZ} d^3k \  \epsilon_n\big({\bf k},
c( {\scriptscriptstyle  {{\bf l}+{\bf l}' \over 2} }) \big)
e^{i {\bf k} \cdot ({\bf l}-{\bf l}')}  
\label{effective Hamiltonian}
\end{equation}
determined by the {\it local band structure} 
$\epsilon_n\big({\bf k},c({\bf r}) \big)$
of the graded crystal \cite{Geller and Kohn}.
Here $\alpha$ is an index labeling the different states 
associated with band $n$, 
and $v$ is the Brillouin zone volume.
The generalized Wannier functions $| a_{n {\bf l}} \rangle $ 
in a nearly periodic
potential are labeled by a band index $n$   
and a lattice vector ${\bf l}$ about which the function 
is localized. They have the same
completeness, orthonormality, and band-diagonality 
properties as the conventional
Wannier functions for strictly periodic 
potentials \cite{GWF}. The use of generalized Wannier functions
makes it possible to construct the stationary 
states (\ref{expansion}) from a {\it single band},
and hence, we label them by the pair of indices $n$ and $\alpha$. 
This is possible as long as $c({\bf r})$ and $U({\bf r})$ are slowly varying.

The dynamic conductivity of an electron in a
state $\psi_{n \alpha}$ may be obtained from
time-dependent perturbation theory. The real part is given by
\begin{eqnarray}
\sigma_1^{ij}(\omega) &=& {\pi e^2 \hbar \over 2 m V }  
{\sum_{n' \alpha '}}'
f^{ij}_{n \alpha , n' \alpha'}     
\bigg[ 
\delta( \epsilon_{n' \alpha'} - \epsilon_{n \alpha} - \hbar \omega) 
+ \delta( \epsilon_{n' \alpha'} - \epsilon_{n \alpha} + \hbar \omega) 
\bigg]                  \nonumber \\     
&+&  { \pi e^2 \over V} \bigg( {1 \over m} \delta^{ij}
- {1 \over m} {\sum_{n' \alpha'}}'
f^{ij}_{n \alpha , n' \alpha'} \bigg) \ \delta (\omega) ,     
\label{real part}
\end{eqnarray}
where 
\begin{equation}
f^{ij}_{n \alpha , n' \alpha'}
\ \equiv \ {2 \over m} { {\rm Re}
\  \langle \psi_{n \alpha} | p^i | \psi_{n' \alpha'} \rangle
\ \langle \psi_{n' \alpha'} | p^j | \psi_{n \alpha} \rangle 
\over  \epsilon_{n' \alpha'} - \epsilon_{n \alpha}  }
\label{oscillator strength}
\end{equation}
is the oscillator strength for an optical transition
between states
$\psi_{n \alpha}$
and
$\psi_{n' \alpha'}$.
Here $e$ and $m$ denote the electron charge and mass, respectively,
and $V$ is the volume of the crystal.
The primed summation in (\ref{real part}) means that the 
$n' \alpha'  = n \alpha$
term is to be excluded.
The real part of $\sigma^{ij}(\omega)$
satisfies the well-known
sum rule
\begin{equation}
\int_0^\infty d\omega \ \sigma_{1}^{ij}(\omega) = { \pi e^2 \over
2 m V } \ \delta^{ij}  ,
\label{conductivity sum rule}
\end{equation}
which reflects the fact that, when driven at a 
high enough frequency, an electron always
responds as though it were free \cite{Mahan}.

We now use (\ref{real part}) 
to review the optical properties
of a uniform crystal that will be 
important for our subsequent developments.
The optical spectrum of a Bloch electron,
$| \varphi_{n {\bf k}},c \rangle  \equiv V^{-{1 \over 2}} 
\sum_{\bf l} e^{i{\bf k} \cdot {\bf l}}
\ | a_{n {\bf l}}^{0} ,c \rangle ,$
consists of a series of absorption and stimulated emission 
lines with frequencies
$[ \epsilon_{n'}({\bf k},c) - \epsilon_n({\bf k},c) ]/ \hbar$, 
where $n' \neq n$,
and with intensities characterized by the oscillator strengths
\begin{equation}
f^{ij}_{nn'}({\bf k},c) \equiv {2 \over m} 
{ {\rm Re} \ p^i_{n n'}({\bf k},c)
\ [ p^j_{n n'}({\bf k},c) ]^* 
\over \epsilon_{n'}({\bf k},c) - \epsilon_n({\bf k},c) },
\label{Bloch oscillator strengths}
\end{equation}
where the
${\bf p}_{n n'}({\bf k},c) \equiv 
\langle \varphi_{n {\bf k}},c | {\bf p} | 
\varphi_{n' {\bf k}},c \rangle$
are momentum matrix elements.
Here $|a_{n{\bf l}}^0 ,c \rangle$ denotes a Wannier 
function for a uniform crystal with
composition $c$.
In a cube of volume $L^3$, the
components of the allowed ${\bf k}$ satisfy $e^{i k L } = 1$. 
The oscillator strengths (\ref{Bloch oscillator strengths}) 
satisfy the sum rule
\begin{equation}
\sum_{n' \neq n} f^{ij}_{n n'}({\bf k},c) = 
\delta^{ij} - {m \over \hbar^2}
{\partial^2 \epsilon_n({\bf k},c) \over \partial k_i \partial k_j } .  
\label{Bloch f sum rule}
\end{equation}
In addition to these finite-frequency spectral lines, 
there is a Drude peak at zero
frequency,
\begin{equation} 
\sigma_1^{ij}(\omega) \sim {\pi e^2 \over \hbar^2 V} 
{\partial^2 \epsilon_n({\bf k},c) \over 
\partial k_i \partial k_j } \ \delta(\omega) . 
\label{Drude peak}
\end{equation}

The conventional analysis described above exploits 
the discrete translational invariance of an infinite
crystal by imposing periodic boundary conditions on a 
finite volume $V$ and by choosing eigenfunctions
with a definite crystal momentum. We shall find it more 
convenient in what follows, however,
to describe the same physical system by instead imposing hard-wall
boundary conditions on a cube of volume $V$, and then taking the 
limit $V \rightarrow \infty$.
In this representation, the eigenfunctions may be taken to be
$| \psi_{n {\bf k}} ,c \rangle \equiv
 V^{-{1 \over 2}} \sum_{\bf l} \sin ({\bf k} \cdot {\bf l})
\ | a_{n{\bf l}}^0 ,c \rangle , $
where the components of the 
allowed ${\bf k}$ satisfy $\sin(k L) = 0$.
For a finite $V$, optical transitions may occur 
between $\psi_{n {\bf k}}$ and any other state
$\psi_{n' {\bf k'}}$ with opposite parity, the 
strongest transitions occuring between
states with ${\bf k}' \approx {\bf k}$. However, 
the widths of the absorption and
emission peaks scale as $1/V^2$, and the optical 
spectrum of a Bloch electron is
recovered in the $V \rightarrow \infty$ limit. 
Because we have chosen hard-wall boundary
conditions, the oscillator strengths satisfy a 
different sum rule, namely
${\sum}'_{n' {\bf k}'} f^{ij}_{n {\bf k}, n' {\bf k}'} = \delta^{ij}$,
where the prime means that the 
$n' {\bf k}' = n {\bf k}$ term is to be excluded.
As expected, this condition implies that the Drude 
peak in (\ref{real part})
vanishes. However, (\ref{conductivity sum rule}) is
still satisfied.

We return now to the case of general 
stationary states $\psi_{n \alpha}$
in a periodic or nearly periodic potential, 
with hard-wall boundary conditions.
We shall consider {\it partial sums} of oscillator strengths
of the form
\begin{equation}
{\sum_{\alpha'}}' f^{ij}_{n \alpha , n' \alpha'} . 
\label{partial sum}
\end{equation}
When $n' \neq n$, (\ref{partial sum}) is the sum of 
interband oscillator strengths
between $\psi_{n \alpha}$ and all states $\psi_{n' \alpha'}$ 
in band $n'$, whereas, when
$n' = n$, it is the sum of all oscillator strengths 
between $\psi_{n \alpha}$ and all
{\it other} states in the same  band. 
In terms of the projection operator 
${\rm P}_{\! n} \equiv \sum_\alpha |\psi_{n \alpha} \rangle  
\langle \psi_{n \alpha} |$,
(\ref{partial sum}) may be written generally as 
\begin{equation}
{\sum_{\alpha' }}' f^{ij}_{n \alpha, n' \alpha' }  
= { \big\langle \psi_{n \alpha} \big| x^i \ \! {\rm P}_{\! n'} \ \! p^j 
- p^j \ \! {\rm P}_{\! n'} \ \! x^i \big| \psi_{n \alpha} \big\rangle
\over i \hbar } .
\label{general sum rule}
\end{equation}
It is clear from (\ref{general sum rule})
that the partial sums depend only on the states 
$\psi_{n \alpha}$ and on the properties
of the crystal in the absence of applied fields, 
through the ${\rm P}_{\! n}$.

We now turn to the evaluation of (\ref{general sum rule}). 
First note that, because $\sum_n {\rm P}_{\! n} = 1$, we have 
${\sum}'_{n' \alpha'} f^{ij}_{n \alpha, n' \alpha'} = \delta^{ij},$
as expected.

Next, we obtain an {\it intraband sum rule} 
from (\ref{general sum rule})
with $n' = n$. A direct calculation yields
\begin{equation}
\sum_{\alpha' \neq \alpha} f^{ij}_{n \alpha, n \alpha' }  
= { \big\langle \psi_{n \alpha} \big| [ x^i_n , p^j_n ] 
 \big| \psi_{n \alpha} \big\rangle
\over i \hbar },
\label{projected commutator}
\end{equation}
where
$O_n \equiv {\rm P}_{\! \! n} O {\rm P}_{\! \! n}$ 
denotes a projected operator.
This commutator of projected 
position and momentum operators is easily 
evaluated in a Wannier
function basis, using  
${\bf p} = (m / i \hbar) [{\bf r},H]$,
where $H$ is the microscopic Hamiltonian.
This leads to the intraband sum rule 
\begin{equation}
\sum_{\alpha' \ne \alpha} \ f^{ij}_{n \alpha , n \alpha'} 
= - {m \over \hbar^2}  
\ \sum_{{\bf l l}'} 
\big( \Phi_{n{\bf l}}^{\alpha} \big)^* 
{\cal H}^n_{\bf l l'}
\big( {\bf l}-{\bf l}' \big)^i
\big( {\bf l}-{\bf l}' \big)^j
\ \Phi_{n{\bf l}'}^{\alpha} .
\label{intraband sum rule}
\end{equation}
The sum of the oscillator strengths for transitions between
a given state 
$\psi_{n \alpha}$ and the other states in the same band,
is therefore proportional to the expectation value of the second moment
of the effective Hamiltonian (\ref{effective Hamiltonian}) in the state
$\psi_{n \alpha}$.
The definition (\ref{effective Hamiltonian}) can also be used 
to show that this second moment is simply
the Fourier component of the local energy band {\it curvature}.
Hence, 
\begin{equation}
\sum_{\alpha' \ne \alpha} \ f^{ij}_{n \alpha , n \alpha'} 
= \sum_{{\bf l l}'} 
\big( \Phi_{n{\bf l}}^{\alpha} \big)^* 
\ \Phi_{n{\bf l}'}^{\alpha} 
\ {1 \over v} \int_{\scriptscriptstyle \rm BZ} 
\ d^3k \ e^{i{\bf k} \cdot ({\bf l}-{\bf l}')} 
\ {m \over \hbar^2} \bigg( {\partial^2 \epsilon_n \big( {\bf k},
c({\scriptscriptstyle {{\bf l}+{\bf l}' \over 2}}) \big) \over
\partial k_i \partial k_j }\bigg) . 
\label{curvature}
\end{equation}
In the special case of a uniform crystal,
$\psi_{n \alpha} \rightarrow \psi_{n {\bf k}}$ and
\begin{equation}
\sum_{{\bf k}' \neq {\bf k}} f^{ij}_{n {\bf k}, n {\bf k}'}
= {m \over \hbar^2} {\partial^2 \epsilon_n({\bf k},c)
\over \partial k_i \partial k_j}.
\end{equation}

The results (\ref{intraband sum rule}) and (\ref{curvature}) 
are valid for any state
$\psi_{n \alpha}$
in the band $n$.
However, if the state 
is near in energy to a band edge, the sum rule 
simplifies to an expectation value of the local inverse
effective mass tensor 
\begin{equation}
\bigg( {1 \over m^*({\bf r}) }\bigg)^{ \! ij}
\equiv \ {1 \over \hbar^2}
\bigg( {\partial^2 \epsilon_n\big(0, c({\bf r}) \big)  
\over \partial k_i \partial k_j } \bigg) .
\label{effective mass}
\end{equation}
We shall assume that the state
$\psi_{n {\alpha}}$
is near the minimum of a simple band and that the minimum
occurs at the center of the Brillouin zone of the local
band structure.
Then the Wannier function amplitudes
$\Phi_{n{\bf l}}^{\alpha}$
vary slowly in space, and a smooth interpolating
function $F_{n \alpha}({\bf r})$ satisfying
$F_{n \alpha}({\bf l}) = \Phi_{n{\bf l}}^\alpha$
may be introduced.
This interpolating function is equivalent to the envelope function
employed in \cite{Geller and Kohn}, where it was shown 
to be an eigenfunction
of an effective mass Hamiltonian with position-dependent
effective mass.
Following the methods of Ref.~\cite{Geller and Kohn},
the intraband sum rule near
a band edge becomes simply
\begin{equation}
\sum_{\alpha' \ne \alpha} \ f^{ij}_{n \alpha , n \alpha'} 
= \ m \ \bigg( F_{n \alpha} \bigg|
\bigg( {1 \over m^*({\bf r}) }\bigg)^{ij}
\bigg| F_{n \alpha} \bigg) ,
\label{band edge intraband f sum rule}
\end{equation}
where
$\big(F \big| O \big| F \big) \equiv
\int d^3r \ F^* O F $
is the conventional envelope function expectation value.
In a uniform crystal, the right-hand side of
(\ref{band edge intraband f sum rule})  
reduces to $m/m^*$.
The sum rule (\ref{band edge intraband f sum rule}) has 
also been derived recently
by Dav\'e and Taylor \cite{Dave and Taylor}.

The intraband sum rule leads immediately to a sum rule for the
${\it intraband}$ contribution to $\sigma_1^{ij}(\omega)$, namely
\begin{equation}
\int_{\rm intra} d\omega \ \sigma_1^{ij}(\omega)  
=  - {\pi e^2 \over 2 V \hbar^2} 
\sum_{{\bf l l}'} 
\big( \Phi_{n{\bf l}}^{\alpha} \big)^* 
{\cal H}_{\bf l l'}
\big( {\bf l}-{\bf l}' \big)^i
\big( {\bf l}-{\bf l}' \big)^j
\ \Phi_{n{\bf l}'}^{\alpha} . 
\label{intraband conductivity sum rule}
\end{equation}
Similarly, for a state near a band edge,
\begin{equation}
\int_{\rm intra} d\omega \ \sigma_{1}^{ij}(\omega)
= {\pi e^2 \over 2 V} \bigg( F_{n \alpha} \bigg| 
\bigg( {1 \over m^*({\bf r})}\bigg)^{ij} 
\bigg| F_{n \alpha} \bigg). 
\end{equation}
These integrations are to be taken from zero to 
the largest intraband frequency.

{\it Interband sum rules} may also be obtained 
from (\ref{general sum rule}) with $n' \neq n$.
Using a Wannier function basis, we find
\begin{equation}
\sum_{\alpha' } \ f^{ij}_{n \alpha , n' \alpha'} 
= \sum_{{\bf l l}'} 
\big( \Phi_{n{\bf l}}^{\alpha} \big)^* 
\ \Phi_{n{\bf l}'}^{\alpha}  
\ {1 \over v} \int_{\scriptscriptstyle \rm BZ} 
d^3k \ f_{nn'}^{ij}\big({\bf k},
c( {\scriptscriptstyle  {{\bf l}+{\bf l}' \over 2} }) \big)
\ e^{i {\bf k} \cdot ({\bf l}-{\bf l}')} ,
\label{interband f sum rule}
\end{equation}
where $f^{ij}_{nn'}({\bf k},c)$ is the conventional Bloch electron
oscillator strength defined in (\ref{Bloch oscillator strengths}).
The sum rule (\ref{interband f sum rule}) shows how 
the Bloch electron oscillator strength
between two 
bands in a uniform crystal is {\it redistributed}
by a composition gradient or applied potential.
In the special case of a uniform crystal, 
(\ref{interband f sum rule}) becomes
$ \sum_{{\bf k}'} f^{ij}_{n {\bf k}, n' {\bf k}'} 
= f^{ij}_{nn'}({\bf k},c). $ 
For a state near in energy to a band edge,
the interband sum rule reduces to
\begin{equation}
\sum_{\alpha'} \ f^{ij}_{n \alpha , n' \alpha'} 
=  \bigg( F_{n \alpha} \bigg|
f^{ij}_{nn'} \big( 0 , c({\bf r}) \big)
\bigg| F_{n \alpha} \bigg) ,
\label{band edge interband f sum rule}
\end{equation}
an extremely simple form indeed.

The sum rules derived here, which apply to uniform and slowly 
graded crystals, complement
the sum rules derived recently for intersubband 
transitions in superlattices \cite{Peters etal}
and quantum wells \cite{Sirtori etal} with abrupt interfaces. 

We now discuss the physical origin of the new sum rules.
For this purpose we shall examine the
stability of {\it double sums} of oscillator strengths of the form 
\begin{equation}
S_{\! n n'}^{ij} \equiv \sum_\alpha \sum_{\alpha'} 
f^{ij}_{n \alpha, n' \alpha'} ,
\label{double sum}
\end{equation}
in the presence of perturbations.
Double sums of this type are commonplace in the theory 
of atomic spectra, where they
characterize the {\it total} optical absorption 
strength between two multiplets $n$ and $n'$, the
individual degenerate or nearly degenerate states in each 
multiplet being labeled by $\alpha$ and
$\alpha'$ respectively. The {\it invariance} of the total 
absorption strength between two
multiplets, under arbitrary unitary transformations among the 
degenerate or nearly degenerate states
in each multiplet, is known as the {\it principle of spectroscopic
stability} \cite{Condon and Shortley}. 
In the atomic physics context, these
unitary transformations usually arise from the 
application of a weak electric or magnetic field.
This principle is not immediately
applicable to our double sum (\ref{double sum}), however, 
because (\ref{double sum}) describes
the total absorption strength between {\it bands} of 
states which are not nearly 
degenerate. Therefore, we shall need to prove a stronger 
version of the stability principle.
By using the identity
$\langle \psi_{n \alpha} | p^i | \psi_{n' \alpha'} \rangle
= i m \big( \epsilon_{n \alpha} - \epsilon_{n' \alpha'} \big)
\langle \psi_{n \alpha} | x^i 
| \psi_{n' \alpha'} \rangle / \hbar,$
valid for bounded systems only, 
(\ref{double sum}) may be written as
\begin{equation}
S_{\! n n'}^{ij} = {1 \over i \hbar} \ \! {\rm Tr}_{n} 
\big( x^i \ \! {\rm P}_{\! n'}
\ \! p^j - p^j \ \! {\rm P}_{\! n'} \ \! x^i \big) ,
\label{partial trace}
\end{equation}
where the partial trace acts in the subspace spanned by band $n$,
and where 
${\rm P}_{\! n} $
is the projection operator for band $n$.

The invariance of the total optical absorption strength 
$S$ under arbitrary unitary 
transformations within the band $n$ and
independently within the band $n'$ is now evident. 
In particular, $S$ will be invariant
under the action of slowly varying perturbations. 
Because $S$ is conserved in going from a periodic potential
to a nearly periodic one, the optical spectrum of the latter is
related to that of the former by an {\it intraband redistribution} of
the transition strengths.
This fact implies the existence of general sum 
rules for the oscillator strengths
of an electron in a nearly periodic potential, 
as demonstrated above.

As an application of our results,
we shall calculate the optical absorption spectrum
of an electron near the bottom of a slowly graded
${\rm Al_c Ga_{1-c} As}$
parabolic quantum well
\cite{Gossard etal}.
In particular, we shall examine the effects of the position-dependent
band structure on the intraband oscillator strengths and selection 
rules. The eigenstates of interest here are near in energy
to the minimum of the local conduction band and hence
may be described by the effective  
Hamiltonian \cite{Bastard etal,Geller and Kohn}
\begin{equation}
H = - {\hbar^2 \over 2} \nabla_i
\bigg( {1 \over m^*({\bf r}) } \bigg)^{\! ij} \! \nabla_j
\ + \ {\cal E}({\bf r}). 
\label{effective mass Hamiltonian}
\end{equation}
Here ${\cal E}({\bf r}) = {\textstyle {1 \over 2}} m^* \omega_0^2 z^2,$ 
where $m^*$
is the electron effective mass in GaAs, and 
$\hbar \omega_0 $
is the energy level spacing at the bottom of the well.
The laser field is assumed to be polarized in the $z$ direction.
The effective mass for an electron at the $\Gamma$ point of
${\rm Al_c Ga_{1-c} As}$
is known to be well-described by a linear interpolation between
the effective mass of
GaAs and  the  
$\Gamma$-point effective mass
of AlAs.
The position-dependent
effective mass in the quantum well may be written as 
$ m^*(z) = m^* 
[ 1 + \eta (z / \ell)^2 ],$
where $\ell^2 \equiv \hbar / m^* \omega_0$,
and where
$\eta $
is a dimensionless quantity characterizing
the relative change in the effective mass over a length
$\ell$.
In an 
${\rm Al_c Ga_{1-c} As}$
parabolic quantum well with $\hbar \omega_0 \approx 1 \ {\rm meV}$,
it can be shown that
$\eta < < 1$.
Thus, the effective-mass gradient 
may be treated perturbatively.
To this end, we shall write (\ref{effective mass Hamiltonian}) as
$ H = H^0 + H^1$,
where
\begin{equation}
 H^1 \ \equiv \ \eta \hbar \omega_0 \bigg( {\textstyle{1 \over 2}}  z^2
{\partial^2 \over \partial z^2 } 
+ z {\partial \over \partial z} \bigg).
\end{equation}
The energies and normalized envelope functions of
$H^0$ 
are given by
$ \epsilon_{j{\bf k}}^0 = ( j + {\textstyle{1 \over 2}} ) \hbar \omega_0
+ \hbar^2 {\bf k}^2 / 2 m^* $
and
$F_{j {\bf k}}^0({\bf r}) = (2^j j! \pi^{1 \over 2} \ell
L^2 )^{-{1 \over 2}} \ \!  e^{i {\bf k} \cdot {\bf r}} 
\ \! e^{-{1 \over 2}(z/\ell)^2}
\ \!  H_j(z / \ell ),$ 
where ${\bf k}$
is a wavevector in the plane of the quantum well, the
$H_j$ are Hermite polynomials,
and where we have used periodic boundary conditions
in the $x$ and $y$ directions.
The envelope functions
determine the actual eigenfunctions of the electron in the $n$th
band through 
$ \psi_{n j {\bf k}}({\bf r}) = \sum_{\bf l} F_{j{\bf k}}({\bf l})
\ a_{n{\bf l}}({\bf r}-{\bf l}) $,
a special case of (\ref{expansion}).

We shall calculate the $zz$ components of the
intraband oscillator strengths, 
\begin{equation}
 f_{j{\bf k},j'{\bf k}'} = {2 m \over \hbar^2}
\big( \epsilon_{j'{\bf k}'} - \epsilon_{j{\bf k}} \big)
\big| \big\langle \psi_{nj{\bf k}} \big| z 
\big| \psi_{nj'{\bf k}'} \big\rangle 
\big|^2 , 
\label{zz oscillator strength}
\end{equation}
to first order in $\eta$.
To this order, the perturbed eigenvalues are
$ \epsilon_{j{\bf k}} = \epsilon_{j{\bf k}}^0 - {1 \over 4} \eta
\big( j^2 + j + {3 \over 2} \big) \hbar \omega_0 $.
The nonvanishing intraband oscillator strengths to order $\eta$ are
$f_{j {\bf k}; j+1, {\bf k}}  = 
(m / m^*) [ j + 1 - \eta ( {\textstyle{1 \over 2}} j^2
+ j + {\textstyle{1 \over 2}} )]$ 
and
$ f_{j {\bf k}; j-1, {\bf k}} = 
(m /m^*) [ - j   + \eta ( {\textstyle{1 \over 2}} j^2 )].$
The selection rules $j \rightarrow j \pm 1$
are therefore unchanged to first order in the effective mass gradient.
However, the oscillator strengths are indeed modified, and
the optical absorption frequencies,
$ \epsilon_{j+1,{\bf k}} - \epsilon_{j,{\bf k}} 
= [ 1 - \eta {\textstyle{1 \over 2}} 
(j+1)] \hbar \omega_0 $,
are decreased.
The sum of the intraband oscillator strengths is 
$(m /m^*) [ 1 - \eta
(j + {\textstyle{1 \over 2}} )],$
which is 
in agreement with (\ref{band edge intraband f sum rule})
to order $\eta$.

It is known that deviations from perfect parabolic confinement
and the existence of a position-dependent effective mass both modify the
optical absorption spectrum of an ideal parabolic quantum
well, by changing the level spacing and oscillator strengths.
Is it possible to separate the effects of these two perturbations,
which are always present in real quantum wells?
Our analysis shows that the {\it integrated} intraband
absorption strength depends on the presence of the 
position-dependent effective mass only, and is therefore a direct 
and unique probe of this subtle band-structure effect.

This work was supported by the NSF through Grants Nos. 
DMR-9308011 and DMR-9403908.
It is a pleasure to thank Walter Kohn for useful discussions
on this subject.

\end{document}